\documentstyle[amssymb,secnumtab]{fbssuppl}

\title{Impact of Few-Body Reactions on Explosive Nucleosynthesis:
The Big-Bang and Supernovae}

\author{Toshitaka Kajino\instnr{1,2,3},
Kaori Otsuki\instnr{1},
Shin-ya Wanajo\instnr{1},
Manabu Orito\instnr{1},
and Grant J. Mathews\instnr{4}}
\instlist{National Astronomical Observatory, Mitaka, Tokyo, 181-8588
\and Department of Astronomy, University of Tokyo, Tokyo, 113-0033
\and Department of Astronomical Science, Graduate
University for Advanced Studies, Tokyo, 181-8588
\and Department of Physics, University of Notre Dame,
Notre Dame, IN 46556}

\begin{document}
\input psfig.sty

\maketitle
\begin{abstract}
We study the few body reactions in explosive nucleosynthesis of 
light-to-heavy mass elements in various astrophysical conditions. 
In neutrino-driven winds of gravitational core collapse of SNeII,
several light nuclei as well as heavy unstable nuclei paly 
significant roles in producing r-process elements.  We discuss 
how the three-body reaction $^4$He$(\alpha n, \gamma)^9$Be,
whose cross section is poorly known, plays very critical role.  
We make note that several other reactions in the r-process are 
relevant to a Big-Bang nucleosynthesis.  We discuss briefly 
a Galactic chemical evolution.
\end{abstract}


\section{Introduction}

Big-Bang nucleosynthesis 
in the early Universe is thought to produce the light elements D and 
$^{3,4}$He with small amounts of $^7$Li, $^9$Be and $^{11}$B.  
Since these abundances are used as unique observables to 
determine the cosmological parameter $\Omega_b$, careful studies of 
the few body reactions of light nuclei are highly desired.  
Stars with various masses provide a variety of production sites 
for $^4$He, $^{12}$C and heavier elements.  Very massive stars  
$\geq 10 M_{\odot}$ culminate their evolution by supernova (SN) explosions 
which are also presumed to be most viable candidate for the still unknown 
astrophysical site of r-process nucleosynthesis.  Even in the 
nucleosynthesis of heavy elements, initial entropy and density 
at the surface of proto-neutron stars are so high that nuclear statistical 
equilibrium (NSE) favors production of abundant light nuclei.  In such explosive 
circumstances few-body reactions play a significant role.

The study of the origin of r-process elements is also critical in
cosmology. 
It is a potentially serious problem that the cosmic age of the expanding
Universe  
derived from cosmological parameters may be shorter than the age of 
the oldest globular clusters.  Since both age estimates
are subject to the uncertain cosmological distance scale, 
an independent method has long been needed.  Thorium, which is a typical 
r-process element and has half-life of 14 Gyr, has recently been detected 
along with other elements in very metal-deficient stars.
If we model the r-process nucleosynthesis in these first-generation
stars, thorium can be used as a cosmochronometer 
completely independent of the uncertain cosmological distance scale.
  
In this article, we first demonstrate theoretically that supernova 
explosions of very massive stars could be a viable site for r-process 
nucleosynthesis.  We study the role of few-body reactions and calculate 
the sensitivity of the r-process yields to the cross section for  
$^4$He$(\alpha n, \gamma)^9$Be, this rate is poorly known experimentally
but is  
expected to affect the result strongly. Next we note
several key reactions in the r-process which are relevant to the Big-Bang 
nucleosynthesis.

\section{Explosive Nucleosynthesis in Supernovae}

Recent mesurements using high-dispersion spectrographs with large 
Telescopes or the Hubble Space Telescope has made it possible
to detect minute amounts of heavy elements in faint metal-deficient 
([Fe/H] $\le$ -2) stars. The discovery of r-process elements in these 
stars has shown that the relative abundance pattern for the mass region 
120 $\le$ A is surprisingly similar to the solar system r-process abundances 
independent of the metallicity of the star
as shown in Fig. 1. 
Here metallicity is defined by 
[Fe/H] = log[N(Fe)/N(H)] - log[N(Fe)/N(H)]$_{\odot}$. It obeys the
approximate relation t/10$^{10}$yr $\sim$ 10$^{[Fe/H]}$. 
The observed similarity strongly suggests that the r-process occurs 
in a single environment which is independent fo progenitor metallicity.
Massive stars with 10$M_{\odot} \le M$ have a short life 
$\sim 10^7$ yr and eventually end up as violent supernova explosions, 
ejecting material into the intersteller medium early on quickly from the
history of the Galaxy.   
However, the iron shell in SNe is excluded from being the 
r-process site because of the observed metallicity independence.  

Hot neutron stars just born in the gravitational core collapse of 
SNeII release most of their energy as neutrinos during the Kelvin-Helmholtz 
cooling phase.  An intense flux of neutrinos heat the material near the
neutron  
star surface and drive matter outflow (neutrino-driven winds).
The entropy in these winds is so high that the NSE favors a plasma
which consists of mainly free nucleons and alpha particles rather than
composite nuclei like iron.  
The equilibrium lepton fraction 
$Y_e$ is determined by a delicate balance between 
$\nu_e + n \rightarrow p + e^-$ and $\bar{\nu}_e + p \rightarrow n + e^+$,
which overcomes the difference of chemical potential between $n$ and $p$,
to reach $Y_e \sim$ 0.45.  R-process nucleosynthesis occurs because there are
plenty of free neutrons at high temperature. 
This is possible only if seed elements are produced in the correct neutron 
to seed ratio before and during the r-process.

Although Woosley et al.~\cite{woosley94} demonstrated a profound possibility 
that the r-process could occur in these winds, several difficulties were 
subsequently identified.  First, independent non relativistic numerical
supernova models~\cite{witti94} have difficulty producing
the required entropy in the bubble S/k $\sim$ 400.  
Relativistic effects may not be enough to increase 
the entropy dramatically~\cite{qian96, cardall97, otsuki00}.  
Second, even should the entropy be high enough, the effects of neutrino 
absorption $\nu_e + n \rightarrow p + e^-$ and 
$\nu_e + A(Z,N) \rightarrow A(Z+1,N-1) + e^-$ 
may decrease the neutron fraction during the nucleosynthesis process.
As a result, a deficiency of free neutrons could prohibit the
r-process~\cite{meyer95}.   
Third, if neutrinos are massive and have approximate mixing parameters, 
energetic $\nu_{\mu}$ and $\nu_{\tau}$ are converted into $\nu_e$ due to 
flavor mixing. This activates the $\nu_e + n \rightarrow p + e^-$ process 
and results in a deficiency of free neutrons.
\begin{center}
\begin{figure}[hbt]
\mbox{\hspace{1cm}}\psfig{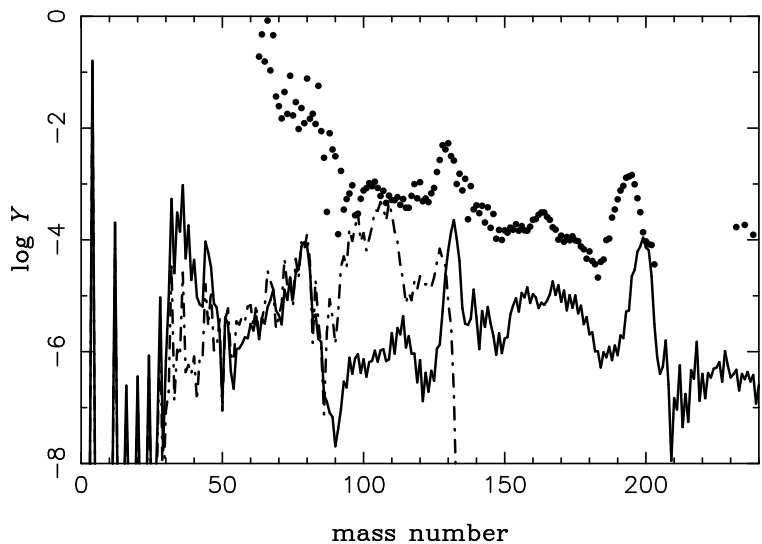}
\caption{
Calculated r-process abundance yields~\cite{otsuki00} in the
neutrino-driven winds as a function of atomic mass number.
Filled circles are the solar system r-process abundances~\cite{kappeler89}, 
shown in arbitrary units. 
}
\end{figure}
\end{center}

In order to resolve these difficulties, we have studied
neutrino-driven winds  
in a Schwarzschild geometry under the  reasonable assumption of spherical 
steady-state flow~\cite{otsuki00}.  Then, we carried out r-process
nucleosynthesis calculations in our wind model~\cite{wanajo99}.
The nuclear reaction network used for light-mass region is shown in
Fig. 2.
The full network consists of $\sim$ 3000 elements including radioactive 
unstable nuclei up to the plutonium isotopes. 

We found~\cite{otsuki00} that the general relativistic effects make expanding 
dynamic time scale $\tau_{dyn}$ much shorter, although the entropy per 
baryon increases by about 40 \% from the Newtonian value of S/k $\sim$ 90.
By simulating many supernova explosions, we have found some interesting
conditions  
which lead to successful r-process nucleosynthesis~\cite{otsuki00, wanajo99}, 
as shown in Fig. 1. The best case is for a relatively 
large neutron-star mass $M > 1.7 M_{\odot}$ and a neutrino luminosity
$L_\nu > 5 \times 10^{51} erg~s^{-1}$. 
\begin{center}
\begin{figure}[hbt]
\psfig{file=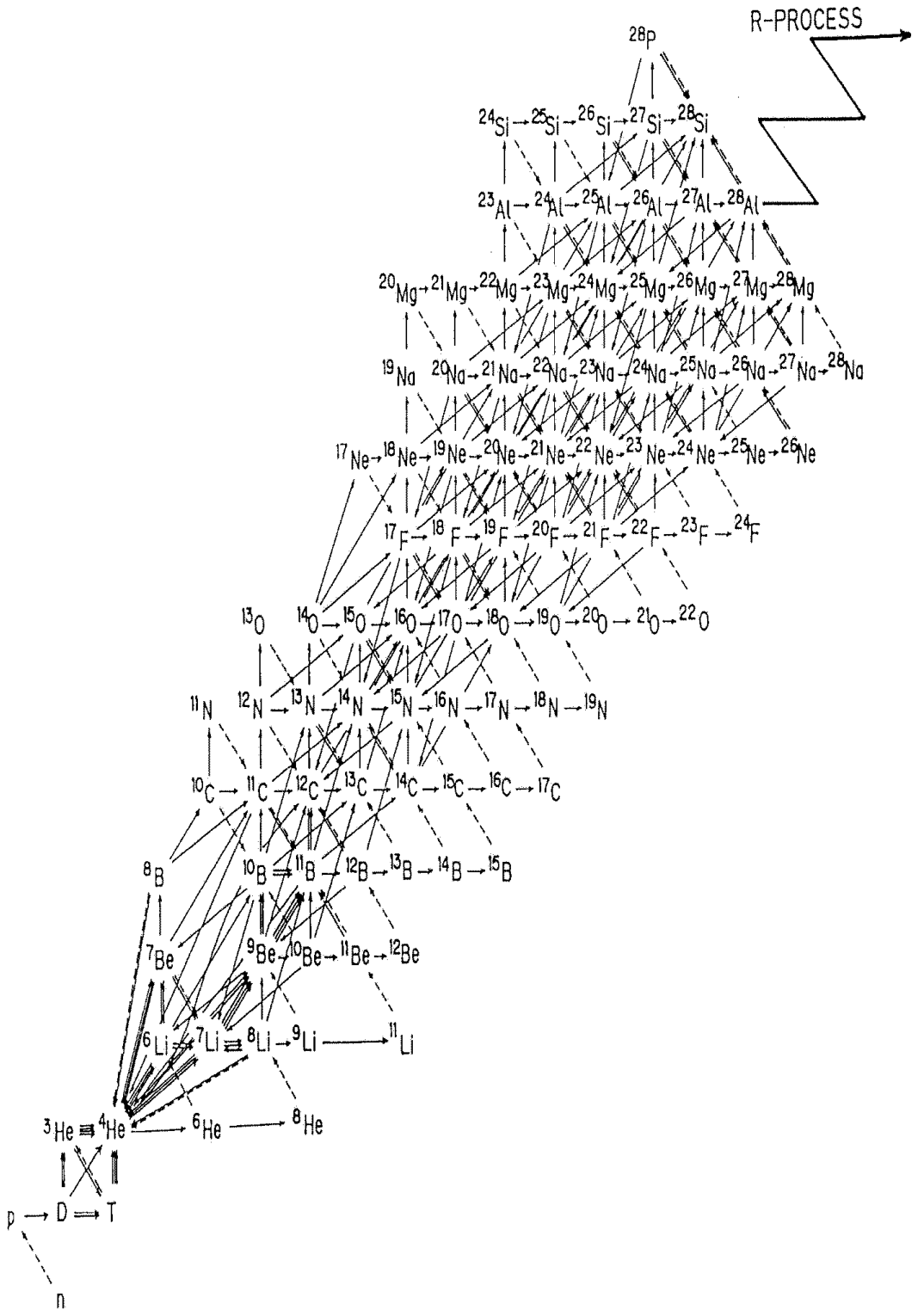,width=11.7cm}
\caption{
Light mass region of the nuclear reaction network used in the calculations.
Solid arrows indicate nuclear reactions in the direction of
positive Q--values. Dashed arrows show beta decays. 
}
\end{figure}
\end{center}

The key to resolve the first difficulty noted above is the short dynamic
time scale  
$\tau_{dyn}\sim$ 10 ms.  As the initial nuclear 
composition of the relativistic plasma consists of neutrons and protons, 
the $\alpha$-burning begins when the plasma temperature cools below 
T $\sim$ 0.5 MeV.  The $^4$He$(\alpha \alpha,\gamma)^{12}$C reaction 
is too slow at this temperature, and alternative nuclear reaction path 
$^4$He$(\alpha n,\gamma)^9$Be$(\alpha,n)^{12}$C triggers explosive 
$\alpha$-burning to produce seed elements with A $\sim$ 100.  
Therefore, the time scale for nuclear reactions is regulated by 
the $^4$He$(\alpha n,\gamma)^9$Be. It is given by 
$\tau_N \equiv \left(\rho_b^2 Y_{\alpha}^2 Y_n \lambda(\alpha \alpha n
\rightarrow ^9{\rm Be}) \right)^{-1}$.
If the neutrino-driven winds fulfill the condition 
$\tau_{dyn} < \tau_N$, then fewer seed nuclei are produced 
during the $\alpha$-process with plenty of free neutrons left over when
the r-process begins at T $\sim$ 0.2 MeV.  The high neutron-to-seed  
ratio, $n/s \sim 100$, leads to appreciable production of r-process 
elements, even for low entropy S $\sim$ 130, producing both the 2nd $(A \sim
130)$ and  
3rd $(A \sim 195)$ abundance peaks and the hill of rare-earth elements 
$(A \sim 165)$ (cf.~Fig.1).

The three body nuclear reaction cross section for $^4$He$(\alpha
n,\gamma)^9$Be is poorly  
determined experimentally.  Because two 
different channels, $^8$Be + n and $^5$He + $\alpha$, contribute to this
process 
it is also a theoretical challenge to understand the reaction mechanism and 
the resonance structure.  We show two calculated results in Fig. 1:
The solid line displays the result obtained by using the cross section
estimated  
by Woosley and Hoffman~\cite{woosley92}, assuming a $^8$Be + n structure
for $^9$Be. 
The cross section estimate may only be good to one order 
of magnitude. Therefore, we also calculated the r-process by multiplying
this cross section  
by a factor of 10 (long-dashed line).  This makes a drastic change in
the r-process abundance  
yields.  Both theoretical and experimental studies of the 
$^4$He$(\alpha n, \gamma)^9$Be reaction are highly desired. 

The specific collision time for neutrino-nucleus interactions is given by
\begin{equation}
\tau_{\nu} \approx 201 \times L_{\nu ,51}^{-1} \times
\left(\frac{\epsilon_{\nu}}{\rm MeV}\right) 
\left(\frac{r}{\rm 100km}\right)^2 
\left(\frac{\langle\sigma_{\nu}\rangle}{\rm 10^{-41}cm^2} \right)^{-1} ms.
\end{equation}
$\tau_{dyn} \approx$ 10 ms $\ll \tau_{\nu} \approx 200$ ms holds
at the $\alpha$-burning site, r $\approx$ 100 km.  This resolves the second 
difficulty:  Because there is not enough time for $\nu$'s to interact with 
n's in such rapidly expanding winds, the neutron fraction is very insensitive 
to the neutrino absorption.

We have recently shown that the neutrino flavor oscillation could destroy the 
r-process condition if the mixing parameters satisfy 0.3 eV$^2 \le \Delta m^2$ 
in our wind model.  Recent experiments of the atmospheric neutrinos and the 
missing solar neutrinos have indicated much smaller $\Delta m^2$, while 
the LSND experiment suggests larger $\Delta m^2$.  
We should wait for more experiments.

\section{Big-Bang Nucleosynthesis and Galactic Chemical Evolution} 

We now found that new possible nuclear reaction paths, which were neglected 
in the previous calculations. In addition to 
$^4$He$(\alpha n,\gamma)^9$Be at relatively late epoch in the r-process,  
there is also $^4$He$(t, \gamma)^7$Li$(\alpha, \gamma)^{11}$B or
$^7$Li$(t, n)^9$Be$(t, n)^{11}$B, followed by $(n, \gamma)$ and beta-decays.  
It is also likely that three-body reactions like 
$^9$Li$(2n, \gamma)$, $^{15}$B$(2n, \gamma)$, $^{24}$O$(2n, \gamma)$, 
$^{27}$F$(2n, \gamma)$, etc., may be important.
This has already been pointed out in the inhomogeneous Big-Bang 
nucleosynthesis model (IBBN)~\cite{kajino90, orito97} 
although the physical conditions are different.

These new paths produce $^9$Be, $^{11}$B and even intermediate-mass nuclei 
12 $\le$ A in the IBBN, the abundances of which are many orders 
of magnitude higher than those of SBBN, as displayed in Fig. 3.
This figure shows also the Galactic chemical evolution.
The rise of the $^9$Be abundance with increasing [Fe/H] arises from spallation 
of Galactic cosmic rays after Galaxy formation.
Each solid curve reaches a different abundance plateau for Big-Bang
nucleosynthesis  
(SBBN or IBBN) in the limit of [Fe/H] = -$\infty$, i.e. t = 0.

The nuclear physics of the few-body reactions of light neutron-rich nuclei
is thus very important not only in studies of r-process nucleosynthesis
but in Big-Bang nucleosynthesis and Galactic chemical evolution as well.
\begin{center}
\begin{figure}[hbt]
\mbox{\hspace{2cm}}\psfig{file=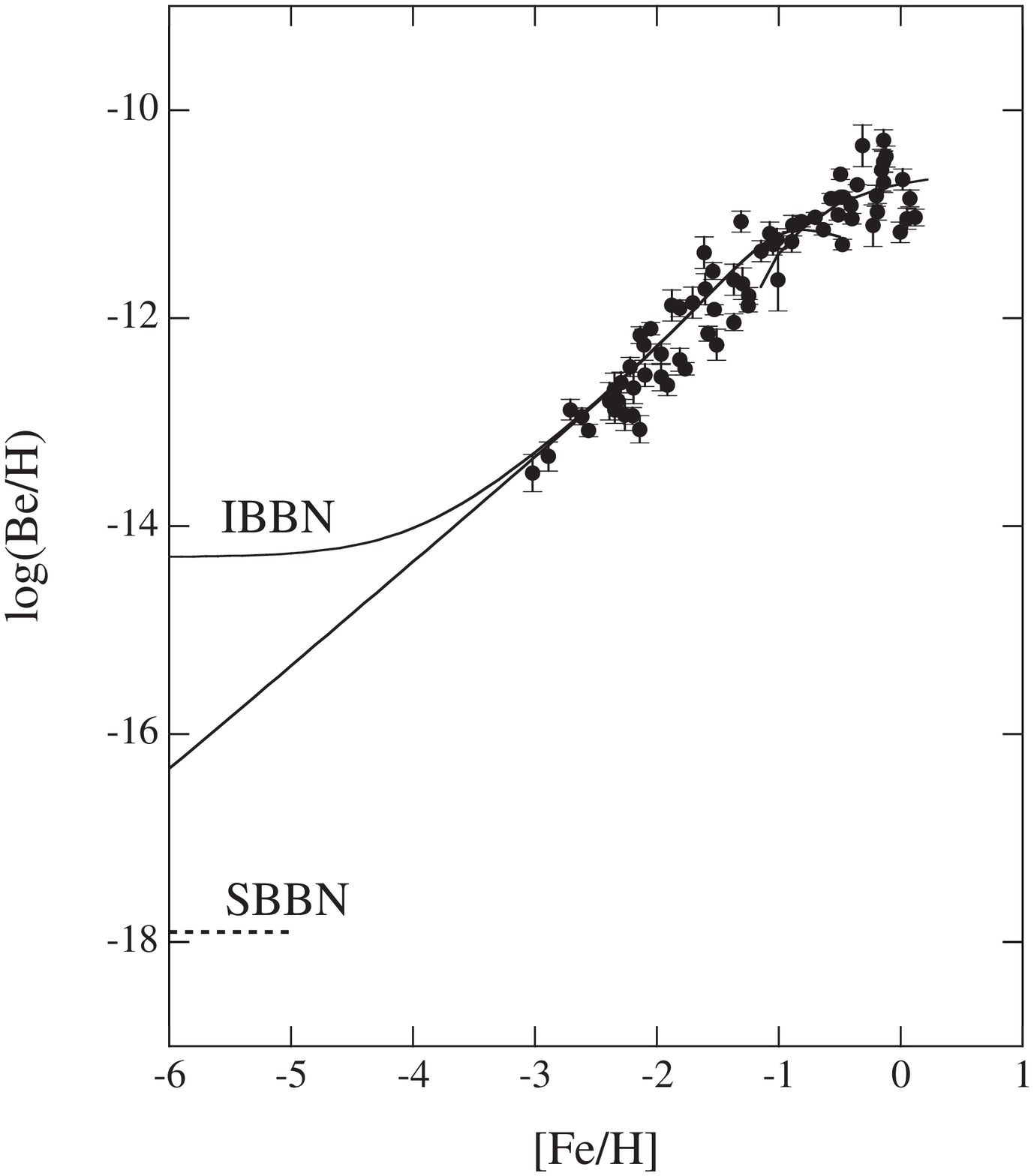,width=7cm}
\caption{ 
Predicted evolution of N($^9$Be)/N(H) vs. [Fe/H] with the standard Big-Bang 
nucleosynthesis model (SBBN; lower dashed plateau) and the inhomogeneous 
Big-Bang nucleosynthesis model (IBBN; upper solid plateau). 
} 
\end{figure}
\end{center}

\SaveFinalPage
\end{document}